\documentclass[11pt,draftcls,onecolumn]{IEEEtran}
\usepackage{graphicx}
\usepackage{cite}

\usepackage{chemarrow}
\usepackage{color}
\usepackage{algorithmic}
\usepackage{algorithm}
\usepackage{clrscode}
\usepackage{amsmath, amssymb, mathrsfs, amsfonts, amsthm}
\usepackage{epsfig, epstopdf}

\usepackage{caption}
\usepackage{subcaption}
\usepackage{stfloats}
\usepackage{enumerate}

\theoremstyle{theorem}
\newtheorem{theorem}{Theorem}
\newtheorem{lemma}{Lemma}

\theoremstyle{remark}

\date{}
\title{ On the Capacity of Level and Type Modulation in Molecular Communication with Ligand Receptors}
\author{\IEEEauthorblockN{Gholamali Aminian\IEEEauthorrefmark{1},
Mahtab Mirmohseni\IEEEauthorrefmark{1}, Masoumeh Nasiri Kenari\IEEEauthorrefmark{1} and Faramarz Fekri \IEEEauthorrefmark{2}}\\
\IEEEauthorblockA{Sharif University of Technology \IEEEauthorrefmark{1} Georgia Institute of Technology \IEEEauthorrefmark{2} }}
\begin{document}
\maketitle
\begin{abstract}
In this paper, we consider the bacterial point-to-point communication problem with one transmitter and one receiver by considering the ligand receptor binding process. The most commonly investigated signalling model, referred to as the Level Scenario (LS), uses one type of a molecule with different concentration levels for signaling. An alternative approach is to employ multiple types of molecules with a single concentration level, referred to as the Type Scenario (TS). We investigate the trade-offs between the two scenarios for the ligand receptor from the capacity point of view. For this purpose, we evaluate the capacity using numerical algorithms. Moreover, we derive an upper  bound on the capacity of the ligand receptor using a Binomial Channel (BIC) model using symmetrized Kullback-Leibler (KL) divergence. A lower bound is also derived when the environment noise is negligible. Finally, we analyse the effect of blocking of a receptor by a molecule of a different type, by proposing a new Markov model in the multiple-type signalling.
\end{abstract}

\section{Introduction}
Molecular communication (MC) has stimulated a great deal of interest because of its potential broad applications. There are different mechanisms for molecular communications, among which diffusion is the most favorable, as it does not require any prior infrastructure. In diffusion-based systems, information is encoded into the concentration, type or releasing time of the molecules. For instance, in \cite{Airfler2011}, an on-off keying modulation is proposed where molecules are released only when the information bit is one. It is shown that if there is no interference from the previous transmission slots, the channel can be modeled by a Z-channel. In \cite{kuran2011 ,tepekule2014}, new modulation techniques based on multiple types of molecules are presented.

Two models for diffusion-based channels have been proposed, namely small and large scales. Diffusion process is viewed as a probabilistic Brownian motion in the small scale model, whereas it is described by deterministic differential equations in the large scale model. In this paper, we concentrate on the \emph{large scale} model which reflects the average effects of diffusion. However, to derive the large-scale diffusion capacity of MC, one has to deal with the reception process at the receiver side. Two reception models are considered for a passive receiver. The first model is a perfect absorber where the receiver absorbs the hitting molecule. The second model, which is more realistic, is the ligand-receptor binding receiver, where the hitting molecule is absorbed by the receptor with some binding probability, \cite{fekri2, atakan2014molecular}. Authors in \cite{pierobon2011noise} interpret the randomness of  ligand binding as noise and employ Markov chains to derive a closed form for it. Ligand receptors are modeled by a Markov chain in \cite{fekri2}, by  a discrete-time Markov model in \cite{CapacityN2}, and by a Binomial channel for a bacterial colony in \cite{fekri1}. The Binomial channel is defined by $p(y|x)={n \choose y} x^y (1-x)^{n-y}$ where the input $x\in [0,1]$, the output $y \in \lbrace 0,1,\ldots,n \rbrace$ and $n$, the number of trials, is a given natural number. Average and peak constraints on the input $x$ may exist. The capacity of this channel without average and peak constraint, for large values of $n$ behaves as follows \cite{xie1997minimax}:
\begin{align}
\frac{1}{2}\log{\frac{n}{2\pi e}}+\log{\pi}
\end{align}
However there is no explicit upper or lower bound on the Binomial channel capacity when $n$ is finite. An algorithm for computing the capacity of Binomial Channel was presented  in \cite{komninakis2001capacity} using convex optimization methods. 

Our main contributions are as follows:
\begin{itemize}
\item We investigate the tradeoffs between two bacterial communication scenarios for ligand receptors: 
(a) multi-type molecular communication with a single concentration level, and (b) single-type molecular communication via multiple concentration levels. At the first glance, scenario (a) introduces a new degrees of freedom  and reduces the intersymbol interference (ISI). However since the number of molecules per type (the power per type) reduces by the increase of the types, we should examine the benefit of using different types of molecules. To make the comparison between scenario (a) and (b), we adopt the model of  \cite{fekri1} in this work.
\item  We derive an upper bound to the capacity of Binomial channel model with given average and peak constraints on the channel input, using KL divergence bound of \cite{aminian} (Theorem \ref{theorem1}). Based on numerical evidence, we believe that this upper bound works well in the low SNR regime (which can occur in MC systems).
\item A lower bound for the Binomial channel with a peak constraint and no environment noise is presented in Lemma~\ref{lemma1}. Based on numerical results, we believe that this lower bound is tight for low peak values.
\item A Markov model for the interactions between different types of molecules near the receptor is presented and the capacity for this model is computed numerically.
\end{itemize}

All logarithms are in base $e$ in this paper. This paper is organized as follows: in Section \ref{sec:model}, we present the system model for  Level and Type signalling scenarios, whose the capacities are discussed in Subsection \ref{Capacity analysis}. In Section \ref{sec:cap_upper}, a new upper bound on the capacity of the Binomial channel is presented by considering peak and average constraints. Section \ref{sec:cap_lower} includes a lower bound on the capacity of the Binomial channel by extending the Z-channel. In Section \ref{sec:block}, the interaction of molecules near the receptor is modeled. Section \ref{Simulation} includes the numerical results.
\section{System Model}\label{sec:model}
\begin{figure}
\begin{center}
\includegraphics[scale=0.5,angle=0]{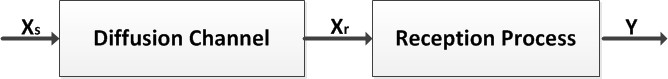}
\end{center}
\caption{The system model}
\label{fig8}
\end{figure}

In this section, we describe two bacterial point-to-point communication scenarios with the ligand receptors, as shown in Fig.~\ref{fig8}. The most commonly investigated model, referred to as the Level scenario (LS), uses one type of molecule with multiple concentration levels for  transmission. An alternative approach is to provide different types of molecules with one concentration level, albeit at a higher cost, referred to as the Type scenario (TS).

\begin{figure}
    \centering
    \begin{subfigure}[b]{0.5\textwidth}
        \centering
        \includegraphics[width=\textwidth]{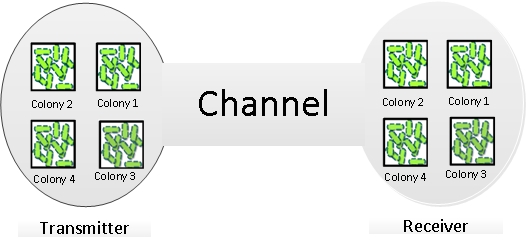}
        \caption{Type scenario (TS) }
        \label{fig2}
    \end{subfigure}
    \hfill
    \begin{subfigure}[b]{0.5\textwidth}
        \centering
        \includegraphics[width=\textwidth]{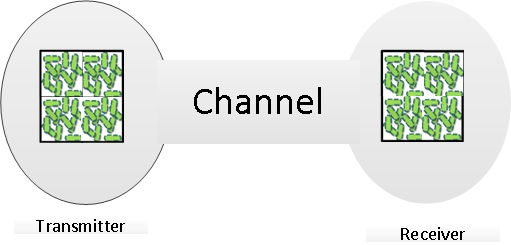}
        \caption{Level scenario (LS)}
        \label{fig3}
    \end{subfigure}
      \caption{Two scenarios: TS and LS}
    \label{fig4}
\end{figure}

\textbf{Level Scenario (LS)}: here the transmitter encodes information at multiple concentration levels to create the codewords. At the transmitter and the receiver, there is only one colony with $n$ bacteria where each bacteria has $N$ receptors; i.e., $nN$ receptors in total. All these $n$ bacteria produce just one type of molecule. This scenario is shown in Fig.~\ref{fig3}.

\textbf{Type Scenario (TS)}: This scenario uses multiple types of molecules at the transmitter and the receiver. We assume the same total number of $n$ bacteria (as in LS) are available which are equally divided into $m$ colonies at both the transmitter and receiver as shown in Fig.\ref{fig2}. As such, each colony has $n/m$ bacteria. Moreover, different colonies at the transmitter produce different types of AHL molecules. Similar to the LS, each bacteria has $N$ receptors. Furthermore, the colonies are synchronized at the transmitter. Therefore, there are ${nN}/{m}$ receptors in total per each colony, i.e., each type of molecule. Each colony can detect its own molecule type and as a result, produces different color Fluorescent Proteins (e.g. GFP,YFP,...) which are used by the receiver to decode the received signal. In addition, we assume all receptors of a colony are independent and sense a common molecule concentration.

Throughout Sections~\ref{sec:model}, \ref{sec:cap_upper} and \ref{sec:cap_lower} we assume that the binding processes of different molecule types are independent. We investigate a more general model in Section~\ref{sec:cap_lower} by taking into account the interaction of different types of molecules at the TS.
In both scenarios, we assume that there is no intersymbol interference (ISI). In other words, we assume those molecules, who do not bind to the receptors in the current time slot, will be degraded to the next time slot and hence will not interfere with molecules from the next slot. This assumption, together with the large-scale diffusion channel property, result in a linear channel. For simplicity, we further assume that no attenuation occurs in the channel. Therefore, the received average concentration $A_r$ is equal to the transmitted average concentration $A_s$.
At the receiver with ligand receptors, the probability of binding at the steady state, is given by  \cite{fekri1}:
\begin{align}\label{bind}
p_b=\frac{A_s}{A_s+\frac{\kappa}{\gamma}}
\end{align}
where $\gamma$ is the input gain and $\kappa$ is the dissociation rate of trapped molecules in the cell receptors. If we consider an environment noise  with average concentration $A_{ne}$, due to the molecules of the same type from other sources, the probability of binding becomes:
\begin{align}\label{bind1}
p_b=\frac{A_s+A_{ne}}{A_s+A_{ne}+\frac{\kappa}{\gamma}}
\end{align}

In LS, we only have one type of molecule and its binding probability is equal to
\begin{align}\label{bindLS}
p_{b_{LS}}=\frac{X_{LS}+A_{ne}^{LS}}{X_{LS}+A_{ne}^{LS}+\frac{\kappa}{\gamma}}, \quad 0 \leq X_{LS} \leq A_s
\end{align}
where $X_{LS}$ is the received average concentration at the receiver and $A_{ne}^{LS}$ is the average concentration of environment noise. On the other hand, in TS, we have different types of molecules; the probability of binding for the $i$th type of molecule is given by
\begin{align}\label{bindTS}
 p_{b_{TS}^i}=\frac{X_i^{TS}+A_{ne}^{i,TS}}{X_i^{TS}+A_{ne}^{i,TS}+\frac{\kappa}{\gamma}}, \quad 0 \leq X_{TS} \leq \frac{A_s}{m}
\end{align}
where $X_i^{TS}$ is the received average concentration of the $i$th type of molecule at the receiver and $A_{ne}^{i,TS}$ is the average environment noise for the $i$th type of molecule. Without loss of generality, we assume all environment noises have equal averages, i.e., $A_{ne}^{i,TS}=A_{ne}^{TS}$. Here, we consider the same $\gamma$ and $\kappa$ for all types of molecules and receptors.

\subsection{Capacity analysis}
\label{Capacity analysis}
 In both scenarios, the output is discrete. Further, we assume an environment noise and consider average and peak signal concentration level constraints. The channel is Binomial as follows:
 \begin{align}\label{Binomialchannel}
 &p(Y=y|X=x)= \\\nonumber
 &{Nn \choose y} f_{p_b}^y(x+A_{ne})  \big ( 1-f_{p_b}(x+A_{ne})\big)^{Nn-y}, \\\nonumber
  &f_{p_b}: [0,\infty] \rightarrow [0,1], y \in \lbrace 0,1, ..., Nn \rbrace
 \end{align}
 Note that for LS we have $A_{ne}=A_{ne}^{LS}$ but for TS we have $A_{ne}=A_{ne}^{TS}$ as the environment noises.
 The function $f_{p_b}(X+A_{ne})$ is the binding probability function where $X$ is the signal concentration level and $A_{ne}$ is the average environment noise. We also assume that the function $f_{p_b}$ is an increasing function and concave. We consider the function $f_{p_b}(X+A_{ne})=\frac{X+A_{ne}}{X+A_{ne}+\frac{\kappa}{\gamma}}$ for ligand receptors.  As we note, when the concentration level increases, the binding probability also increases.\\
In LS, we have a single colony with input $X$ and output $Y$, with the following peak and average concentration constraints:
 \begin{align}\label{LScons}
 0 \leq X \leq  A_s\\
 E[X]\leq \alpha
 \end{align}
Then we find the capacity for LS as
 \begin{align}\label{LScap}
 C_{LS}=\max_{\substack{p(x),\\ 0 \leq X\leq A_s,\quad E[X] \leq \alpha}}I(X;Y), \quad Y \in \lbrace 0,1, ..., Nn \rbrace.
 \end{align}

In TS, we use $X_i$ to denote the input of the $i$th colony to the channel and $Y_i$ to denote the output of the $i$th colony at the receiver. The constraints for TS are as follows
\begin{align}\label{TScons}
0 \leq X_i \leq \frac{A_s}{m} \quad i=1,...,m,\\
E[X_i] \leq \frac{\alpha}{m}, \quad i=1,...,m.
\end{align}
  Hence, the capacity can be written as
\begin{align}\label{TScap}
C_{TS}&= \max_{p(x_1,x_2,...,x_m)}I(X_1,...X_m;Y_1,...,Y_m)\\\nonumber
& =m\times \max_{\substack{p(x),\\ 0 \leq X_i\leq \frac{A_s}{m}, E[X_i] \leq \frac{\alpha}{m} } }I(X_i;Y_i),\quad \quad Y_i \in \lbrace 0,1, ..., \frac{nN}{m} \rbrace.
\end{align}
For a fair comparison of the $C_{LS}$ with $C_{TS}$, we consider $A_{ne}^{LS}=A_{ne}^{TS}$.

\section{Capacity Upper Bound}\label{sec:cap_upper}
There is no closed form for the Binomial channel capacity. As such, for the first time, we propose an upper bound for the Binomial channel at the low SNR regime by considering power and peak constraints, i.e., $E[X] \leq \alpha ,~ 0 \leq X \leq A$ respectively. The Binomial channel is defined as follows:
\begin{align}\label{binomchan}
&P(Y=y|X=x)= \\\nonumber
&{Nn \choose y} f_{p_b}^y(x+A_{ne}) \big(1-f_{p_b}(x+A_{ne})\big)^{Nn-y},\\\nonumber
 &f_{p_b}: [0,\infty] \rightarrow [0,1],y \in \lbrace 0,1, ..., Nn \rbrace
\end{align}
$p(y|x)$ is binomial distribution, we have $\sum_y yp(y|x)=Nnf_{p_b}(x)$.

We introduce a new upper bound on the capacity of the Binomial channel based on the symmetrized Kullback-Leibler (KL) divergence, referred as KL upper bound in \cite{aminian}. We first explain the KL upper bound briefly. Let $D_{\mathsf{sym}}(p\|q)=D(p\|q)+D(q\|p)$. Then,
\begin{align}\label{KL}
\mathcal{U}(p(y|x))&=\max_{p(x)}D_{\mathsf{sym}}(p(x,y)\|p(x)p(y))\\\nonumber
&\geq \max_{p(x)}I(X;Y)=C(p(y|x)).
\end{align}
The KL  $\mathcal{U}(p(y|x))$ is always an upper bound on the capacity. It is straightforward to show that
\begin{align}\label{KL1}
&D_{\mathsf{sym}}\big(p(x,y)\|p(x)p(y)\big)=\\\nonumber
&\mathbb{E}_{p(x,y)}\log p(Y|X) - \mathbb{E}_{p(x)p(y)}\log p(Y|X)
\end{align}

Now, we state our upper bound in the following theorem. The proof is provided in Appendix \ref{AppendixProoftheorem1}.
\begin{theorem}\label{theorem1}
Consider a point to point Binomial channel $P(Y=y|X=x)= {Nn \choose y} f_{p_b}^y(x+A_{ne}) (1-f_{p_b}(x+A_{ne}))^{Nn-y}$, and any input probability mass function (p.m.f) $p(x)$ where $f_{p_b}: [0,\infty] \rightarrow [0,1],y \in \lbrace 0,1, ..., Nn \rbrace$. Then, for any input distribution the symmetrized KL divergence upper bound has the following explicit formula:
\begin{align}
I(X;Y)&\leq \mathcal{U}(p(x,y))\\\nonumber
&=Nn\mathsf{Cov} (f_{p_b}(X+A_{ne}), \log (\frac{f_{p_b}(X+A_{ne})}{(1-f_{p_b}(X+A_{ne}))})),
\end{align}
where $\mathsf{Cov}(X,Y)=\mathbb{E}[XY]-\mathbb{E}[X]\mathbb{E}[Y]$. Furthermore, for a Binomial channel with average intensity constraint $\alpha$ and peak constraint $\mathsf{A}$ we have
\begin{align*}\label{upperbound}
&\mathcal{U}_{\mathsf{Binomial}}(p(y|x)):=\max_{\substack{p(x):\\ E[X]=\alpha, 0 \leq X \leq \mathsf{A}}}\mathcal{U}(p(x,y))\\\nonumber
&=nN\begin{cases}\frac{f_{p_b}(\alpha+A_{ne})}{f_{p_b}(A+A_{ne})}[f_{p_b}(A+A_{ne})-f_{p_b}(\alpha+A_{ne})]E,&\textit{if}\quad(*)
\\\nonumber
\frac{f_{p_b}(A+A_{ne})}{4}E,&\textit{if}\quad(**)
\end{cases}
\end{align*}
where $E=\log{\frac{f_{p_b}(A+A_{ne})(1-f_{p_b}(A_{ne}))}{f_{p_b}(A_{ne})(1-f_{p_b}(A+A_{ne}))}}$, $(*)=f_{p_b}(\alpha+A_{ne})< \frac{f_{p_b}(A+A_{ne})}{2}$, and $(**)=f_{p_b}(\alpha+A_{ne})\geq \frac{f_{p_b}(A+A_{ne})}{2}$ .
Hence,
\begin{align}
C=\max_{\substack{p(x):\\ E[X]=\alpha,~~ 0 \leq X \leq \mathsf{A}}}I(X;Y)\leq \mathcal{U}_{\mathsf{Binomial}}(p(y|x)).
\end{align}
\end{theorem}
We compute this KL upper bound numerically in Section~\ref{Simulation}. Based on numerical evidence, we believe that this upper bound works well for Binomial channels (such as MC channels) with low capacity.
\section{Capacity Lower Bound}\label{sec:cap_lower}
In this section, we compute a lower bound for the Binomial channel when the environment noise is negligible, by assuming a binary input, while in the previous section, continuous input was assumed. Further, we do not consider the average constraint. We compute a closed form formula for the lower bound. As an example, the transition probabilities of the Binomial channel with $A_{ne}=0$, $n=2$, and $N=2$ is shown in Fig.~\ref{fig6}. The proof of the following lower bound is provided in Appendix \ref{AppendixProoflemma1}.

\begin{figure}
\begin{center}
\includegraphics[scale=0.5,angle=0]{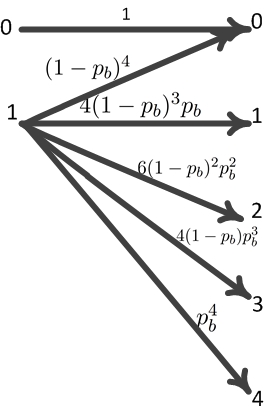}
\end{center}
\caption{Binomial Channel, $n=2, \quad N=2$}
\label{fig6}
\end{figure}

\begin{lemma}\label{lemma1}
Consider a point to point Binomial channel $P(Y=y|X=x)= {Nn \choose y} f_{p_b}^y(x+A_{ne}) (1-f_{p_b}(x+A_{ne}))^{Nn-y}$, and any input p.m.f. $p(x)$, in which $A_{ne}=0$ and $x \in \lbrace 0, A_s \rbrace$. A lower bound on the capacity of this channel is obtained as:
\begin{align}\label{Lowerbound}
C_{lower}=H(\frac{1}{1+e^{g(p_b)}})-\frac{g(p_b)}{1+e^{g(p_b)}}
\end{align}
where $p_b=\frac{A_s}{A_s+\frac{\kappa}{\gamma}}$, $g(p_b)=\frac{H((1-p_b)^{nN})}{1-(1-p_b)^{nN}}$ and $H(p)=-p\log p -(1-p) \log (1-p)$.
The capacity of this channel is a lower bound to the Binomial channel capacity without the energy constraint, when the environment noise is zero.
\end{lemma}
If we consider $Nn=1$ then the channel would reduce to a Z-channel.

\section{Blocking of Molecule near Receptor}\label{sec:block}
\begin{figure}
\begin{center}
\includegraphics[scale=0.5,angle=0]{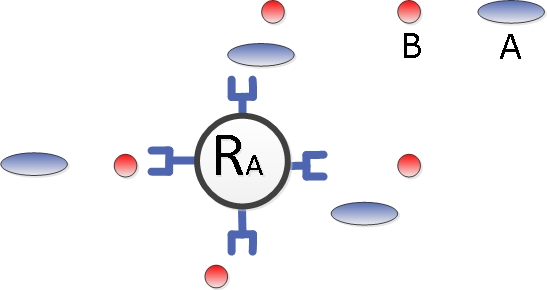}
\end{center}
\caption{Blocking of $R_A$}
\label{fig20}
\end{figure}

In the previous sections, for the TS scenario we assumed orthogonal parallel channels for different types of molecules where there is no interference between different types of molecules (i.e. no blocking of a receptor by molecules of another type). However, it is expected that when there are different types of molecules, they may interfere with each other. In other words, one type of molecule may block another type of molecule from binding to its receptor pair. For example, consider $m=2$ with two types of molecule, $A,B$ and their corresponding receptors as $R_A,R_B$ respectively. The molecule type $A$ near $R_B$ may prevent the molecule type $B$ from binding to $R_B$ and vice versa (see Fig.~\ref{fig20}).

\begin{figure}
    \centering
    \begin{subfigure}[b]{0.5\textwidth}
        \centering
        \includegraphics[width=\textwidth]{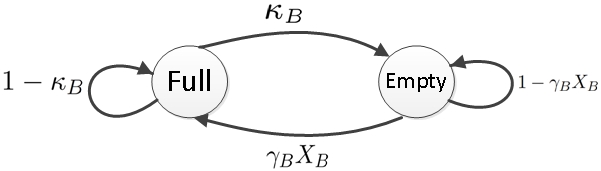}
        \caption{Markov Model with no Blocking for Receptor type $B$ }
        \label{fig12}
    \end{subfigure}
    \hfill
    \begin{subfigure}[b]{0.5\textwidth}
        \centering
        \includegraphics[width=\textwidth]{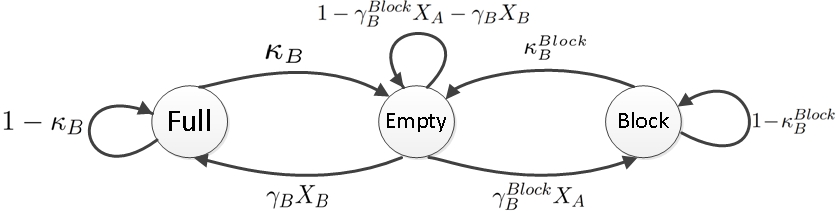}
        \caption{Markov Model with blocking  for Receptor type $B$}
        \label{fig11}
    \end{subfigure}
      \caption{Two Markov Models for Receptor type $B$}
    \label{fig13}
\end{figure}

Assume  that $X_A$ and $X_B$ are the received average concentrations of types $A$ and $B$, respectively. The main reaction kinetics, for binding the molecule type $B$ to its receptor may be modeled as \cite{atakan2014molecular}:
\begin{align}\label{kinetic}
X_B+R_B \underset{\kappa_{B}}{\overset {\gamma_{B}}{\rightleftharpoons}} XR_B,
\end{align}
where $\gamma_{B}\geq 0$ is the association rate of molecules type $B$ with receptors of type $B$ and $\kappa_{B} \geq 0$ is the dissociation rate of $XR_B$ complex. Now, we can characterize the blocking for the receptor type $B$, similar to the reaction kinetics formulas, by:
\begin{align}\label{blockingeq}
&X_A+R_B \xrightarrow{\gamma_{B}^{Block,A}} R_B^{Block,A},\\\nonumber
& X_A+R_B \xleftarrow{\kappa_{B}^{Block,A}} R_B^{Block,A}
\end{align}
where $\gamma_{B}^{Block,A} \geq 0$ is the blocking rate of $R_B$ by molecule type $A$ and $\kappa_{B}^{Block,A}$ is the unblocking rate of $R_B^{Block,A}$ (which $R_B$ was blocked by molecule type $A$). If we do not take the blocking into account, then we have a reaction kinetic for each type of receptor to its type of molecule. As in \cite{atakan2014molecular}, we may define a Markov model for the no blocking case based on (\ref{kinetic}), as shown in Fig.~\ref{fig12} for $m=2$. Likewise, according to (\ref{blockingeq}) and similar to (\ref{kinetic}), we propose a Markov model for the blocking as shown in Fig.~\ref{fig11}.

We consider three states. The full state is when the receptor binds to its type, the empty state when the receptor is free, and the block state when the receptor is blocked with a different type of molecule. The steady state behavior of the system reaction formula can be obtained as (see Fig.~\ref{fig12}):
\begin{align}\label{maineq}
p_b=p_{Full}=\frac{X_B}{X_B+\frac{\kappa_B}{\gamma_B}}
\end{align}

\begin{figure*}
    \centering
    \begin{subfigure}[b]{0.6\textwidth}
        \centering
        \includegraphics[width=\textwidth]{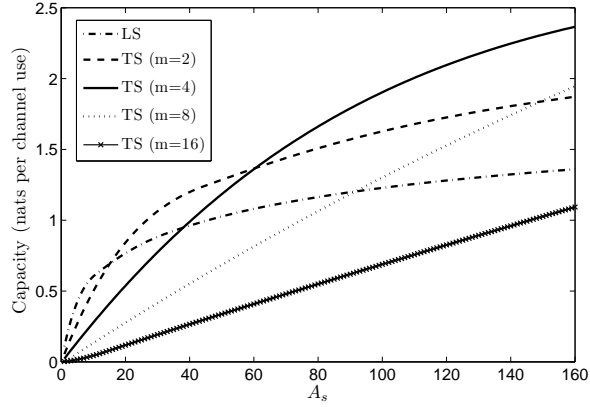}
        \caption{$A_{ne}^{LS}=A_{ne}^{TS}=0$ }
        \label{fig7}
    \end{subfigure}
    \hfill
    \begin{subfigure}[b]{0.6\textwidth}
        \centering
        \includegraphics[width=\textwidth]{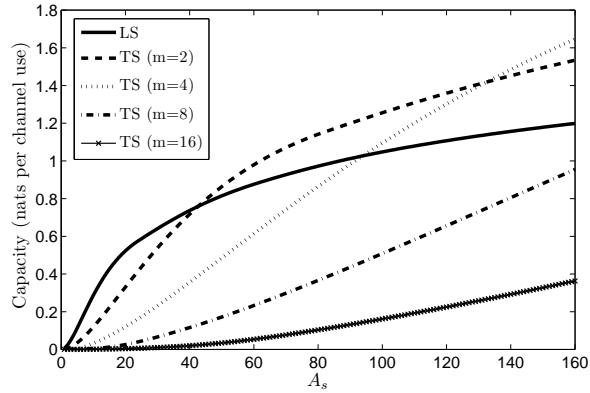}
        \caption{$A_{ne}^{LS}=A_{ne}^{TS}=5$ }
        \label{fig10}
    \end{subfigure}
      \caption{Capacity of LS and TS for $\alpha_{LS}=\frac{A_s}{2m}$ and $ \alpha_{TS}=\frac{A_s}{2}$.}
    \label{fig21}
\end{figure*}

Solving the chain for the blocking case, we have the following probability of binding and blocking:
\begin{align}\label{Blockequa}
&p_b=p_{Full}=\frac{\frac{\gamma_B}{\kappa_B} X_B}{\frac{\gamma_B}{\kappa_B} X_B + \frac{\gamma_B^{Block,A}}{\kappa_B^{Block,A}} X_A+1},\\
&p_{Block}=\frac{\frac{\gamma_B^{Block,A}}{\kappa_B^{Block,A}} X_A}{\frac{\gamma_B}{\kappa_B} X_B + \frac{\gamma_B^{Block,A}}{\kappa_B^{Block,A}} X_A+1}.
\end{align}
If we increase one type of molecule, the probability of binding for another type is decreased as expected. This model can be extended for $m>2$ via,
\begin{align}
&p_b^i=p_{Full}^i=\frac{\frac{\gamma_i}{\kappa_i} X_i}{\frac{\gamma_i}{\kappa_i} X_i + \frac{\gamma_i^{Block}}{\kappa_i^{Block}} \sum_{j=1,j \neq i}^m X_j + 1 } \\
&p_{Block}^i=\frac{\frac{\gamma_i^{Block}}{\kappa_i^{Block}} \sum_{j=1,j \neq i}^m X_j}{\frac{\gamma_i}{\kappa_i} X_i + \frac{\gamma_i^{Block}}{\kappa_i^{Block}} \sum_{j=1,j \neq i}^m X_j + 1 }
\end{align}
where $p_b^i$ and $p_{Block}^i$ are the probability of binding for the $i$th molecule to its receptor and the blocking probability of the $i$th type of the receptor by the another type, respectively. We assume the same blocking and unblocking rates for the $i$th receptor, which are defined by $\gamma_i^{Block}$ and $\kappa_i^{Block}$ respectively. It is also possible to consider the environment noise $A_{ne}$ for the binding and blocking probabilities.
By considering the probability of binding when there is blocking, this channel is a multi-input multi-output Binomial channel, whose capacity is defined as follow:
\begin{align}\label{Mimobino}
&C_{TS}=\max_{\substack{p(x_1,x_2,...,x_m),\\ 0 \leq X_i\leq A_s, E[X_i] \leq \alpha}} I(X_1,...X_m;Y_1,...,Y_m),\\\nonumber
& Y_i \in \lbrace 0,1, ..., \frac{nN}{m} \rbrace\\\nonumber
&P(Y_i=y_i|X_1=x_1,...,X_m=x_m)=\\\nonumber
&{\frac{nN}{m} \choose y_i} f_{p_b^i}^{y_i}(x_1,...,x_m,A_{ne}) (1-f_{p_b^i}(x_1,...,x_m,A_{ne}))^{(\frac{nN}{m}-y_i)}
\end{align}
where $f_{p_b^i}(x_1,...,x_m,A_{ne})=p_b^i$ is the probability of binding when the blocking is considered.
\section{Simulation} \label{Simulation}
In this section, we evaluate the rates of TS scenario given in equation \eqref{TScap}, and the LS scenario given in equation \eqref{LScap},  using Blahut-Arimoto algorithm (BA) \cite{blahut}. We consider $n=16,N=5,\gamma=0.0004$ and $\kappa=0.1$.

\begin{figure}[h!]
\begin{center}
\includegraphics[scale=0.5,angle=0]{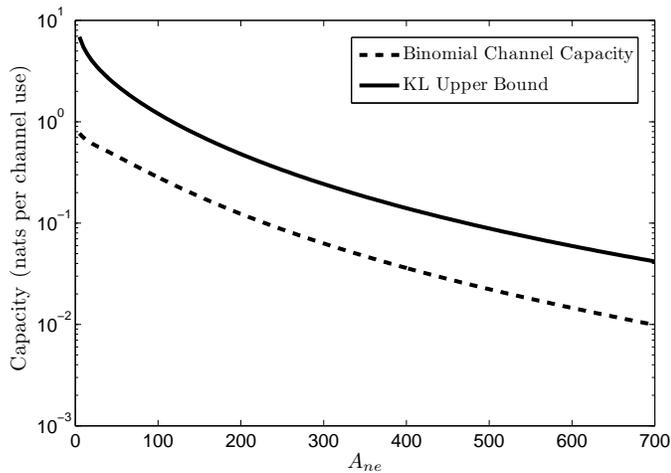}
\end{center}
\caption{Capacity and KL upper bound in terms of $A_{ne}$ for the Binomial channel with $A_s=80$ and $\alpha=40$.}
\label{fig9}
\end{figure}

\begin{figure}[h!]
\begin{center}
\includegraphics[scale=0.5,angle=0]{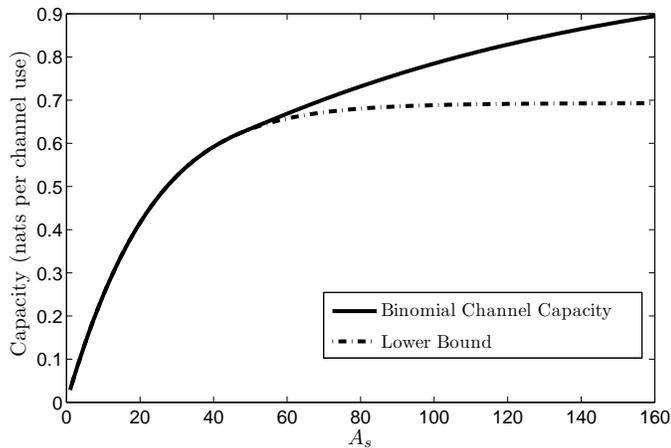}
\end{center}
\caption{Capacity and Lower Bound in terms of $A_s$ for the Binomial channel with $n=4$ and $N=5$.}
\label{fig22}
\end{figure}

\begin{figure}[h!]
\begin{center}
\includegraphics[scale=0.5,angle=0]{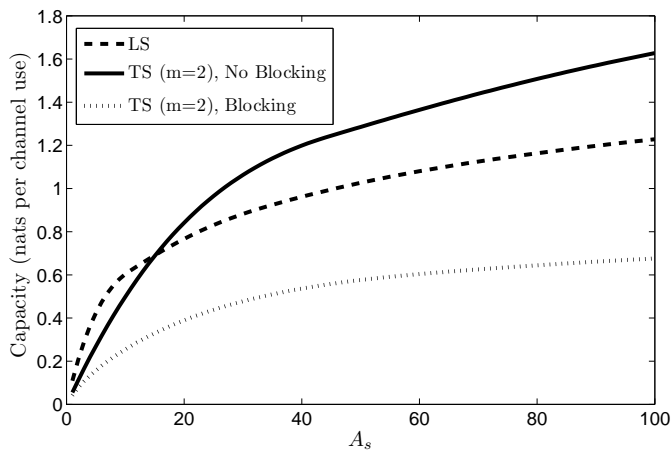}
\end{center}
\caption{Capacity for TS by considering blocking and without blocking and LS for $\alpha_{LS}=\frac{A_s}{2m}$ and $ \alpha_{TS}=\frac{A_s}{2}$, for $A_{ne}=0$.}
\label{fig24}
\end{figure}

Fig.~\ref{fig7} shows the capacity for LS and TS, for $m=2,4,8,16$ when $A_{ne}^{LS}=A_{ne}^{TS}=0$. It is seen that increasing the number of molecule types, $m$, from 1 improves the performance (for fixed $A_s$), which is expected due to the parallel transmission of the molecules. However, if we continue to increase $m$, and accordingly decrease the number of bacteria in each colony to ${n}/{m}$, the performance degrades. The reason is that decreasing the concentration level of TS in equation \eqref{TScons} decreases the binding probability. Hence, there is an optimal $m$. For example, we can see that, for $A_s=80$, this optimal value lies between $m=4$ and $m=8$. This means that for $A_s=80$, at $m=2,4$ the TS capacity is higher than the LS, but for $m=8,16$, it is lower than the LS. Similar conclusions can be made from Fig.~\ref{fig10} in the presence of the environment noise $A_{ne}^{LS}=A_{ne}^{TS}=5$ .
Our proposed KL upper bound, in (\ref{upperbound}), is depicted in Fig.~\ref{fig9}, where the gap between the capacity and the upper bound decreases as the environment noise increases. It can be observed that the distance between the KL upper bound and the capacity is constant in logarithm.
\\
 The lower bound in (\ref{Lowerbound}) along with the capacity are shown in Fig.~\ref{fig22}. For small values of $A_s$, our lower bound is tight which means the binary distribution is capacity achieving distribution for small values of $A_s$.
Fig.~\ref{fig24} shows the effect of blocking by showing the capacity of LS and TS for $m=2$. We assumed $\gamma_1^{Block}=\gamma_2^{Block}=0.0005$, $\kappa_1^{Block}=\kappa_2^{Block}=0.01$, $\gamma_1=\gamma_2=0.0004$ and $\kappa_1=\kappa_2=0.1$. As illustrated, blocking decreases the capacity of TS. In this case, the loss due to blocking is even more than the improvement due to the use of multiple molecule types.

\section{Conclusion}
In this paper, we first investigated capacity performance of type and level scenarios. Next, we derived a new upper bound for the capacity of the Binomial channel at low SNR-regime based on the KL-divergence bound as well as a lower bound. Next, blocking was modeled as a Markov process and the probabilities of binding and blocking were derived. As expected and confirmed by simulations, the blocking would decrease the capacity of type scenario.

\section{Acknowledgements}
The authors would like to thank Dr. Amin Ghohari, for his helpful comments.

\bibliographystyle{ieeetr}
\bibliography{reftest}

\appendices

\section{Proof of Theorem \ref{theorem1}}
\label{AppendixProoftheorem1}
We find KL upper bound for Binomial channel as follows:
\begin{align*}
I(X;Y) &\leq \sum_{x,y} [p(x,y)-p(x)p(y)] \log{p(y|x)}\\
&=\sum_{x,y} [p(x,y) - p(x)p(y)]\\
&\quad \log \big  ( {{Nn \choose y} f_{p_b}^y(x+A_{ne}) (1-f_{p_b}(x+A_{ne}))^{Nn-y}}\big )\\
&=\sum_{x,y} [p(x,y) - p(x)p(y)][\log{{Nn \choose y}}\\
&\quad +y\log{f_{p_b}(x+A_{ne})}\\
&\quad +(Nn-y)\log{(1-f_{p_b}(x+A_{ne}))}]\\
&=E_{p(x,y)}[\log{{Nn \choose y}}]\\
&\quad-E_{p(x)p(y)}[\log{{Nn \choose y}}]\\
& \quad +E_{p(x,y)}[y\log{f_{p_b}(x+A_{ne})}]\\
&\quad -E_{p(x)p(y)}[y\log{f_{p_b}(x+A_{ne})}] \\
& \quad +E_{p(x,y)}[(Nn-y)\log{(1-f_{p_b}(x+A_{ne}))}]\\
&\quad -E_{p(x)p(y)}[(Nn-y)\log{(1-f_{p_b}(x+A_{ne}))}]
\end{align*}
\begin{align*}
&=E_{p(x,y)}[y\log{f_{p_b}(x+A_{ne})}]\\
&\quad-E_{p(x)p(y)}[y\log{f_{p_b}(x+A_{ne})}] \\
& \quad -\big [E_{p(x,y)}[y\log{(1-f_{p_b}(x+A_{ne}))}]\\
&\quad -E_{p(x)p(y)}[y\log{(1-f_{p_b}(x+A_{ne}))}]\big]\\
&=E_{p(x,y)}[y \log{\frac{f_{p_b}(x+A_{ne})}{1-f_{p_b}(x+A_{ne})}}]\\
&\quad-E_{p(x)p(y)}[y \log{\frac{f_{p_b}(x+A_{ne})}{1-f_{p_b}(x+A_{ne})}}]\\
&=\sum_{x}\big((\sum_y y p(y|x))\log{\frac{f_{p_b}(x+A_{ne})}{1-f_{p_b}(x+A_{ne})}}p(x)\big)\\
&\quad-\big(\sum_{x}(\sum_y y p(y|x))p(x)\big)\big(\sum_x \log{\frac{f_{p_b}(x+A_{ne})}{1-f_{p_b}(x+A_{ne})}}p(x)\big)\\
&=E[Nnf_{p_b}(x+A_{ne})\log{(\frac{f_{p_b}(x+A_{ne})}{1-f_{p_b}(x+A_{ne})})}]\\
&\quad -E[Nnf_{p_b}(x+A_{ne})]E[\log(\frac{f_{p_b}(x+A_{ne})}{1-f_{p_b}(x+A_{ne})})]\\
&=Nn\texttt{Cov} (f_{p_b}(X+A_{ne}), \log (\frac{f_{p_b}(X+A_{ne})}{(1-f_{p_b}(X+A_{ne}))})).
\end{align*}
As mentioned earlier, $f_{p_b}(X+A_{ne})$ is an increasing function. Hence the $\texttt{Cov} (f_{p_b}(X+A_{ne}), \log (\frac{f_{p_b}(X+A_{ne})}{(1-f_{p_b}(X+A_{ne}))}))\geq 0$. A further observation is that
$$ C \leq \max_{p(x)} Nn\texttt{Cov} (f_{p_b}(X+A_{ne}), \log (\frac{f_{p_b}(X+A_{ne})}{(1-f_{p_b}(X+A_{ne}))}))$$ is always achievable with a binary rv $X$. We consider two points, $x_1$ and $x_2$ with probabilities $p_1,p_2$. We have
\begin{align*}
&\max_{p(x)} \texttt{Cov} (f_{p_b}(X+A_{ne}), \log (F) )\\
&=\max_{\substack{p(x),\\ E(f_{p_b}(X+A_{ne}))\leq \alpha^\prime, 0 \leq X \leq A}}E[f_{p_b}(X+A_{ne})\log{F}]-\\
& \qquad E[f_{p_b}(X+A_{ne})]E[\log{F}]\\
&=\max_{\substack{p(x),\\ E(f_{p_b}(X+A_{ne}))\leq \alpha^\prime, 0 \leq X \leq A}}E[(f_{p_b}(X+A_{ne})-\\
&\qquad E[f_{p_b}(X+A_{ne})])\log{F}]
\end{align*}
where $F=\frac{f_{p_b}(X+A_{ne})}{1-f_{p_b}(X+A_{ne})}$. Now, based on the analysis in \cite[Appendix C]{aminian}, the optimal distribution is given by $p(x)=\frac{\alpha^\prime}{f_{p_b}(A+A_{ne})}\delta(x-A)+(1-\frac{\alpha^\prime}{f_{p_b}(A+A_{ne})})\delta(x)$ and the upper bound is obtained as:
\begin{align*}
\max_{\alpha^\prime \leq f_{p_b}(\alpha+A_{ne})}\frac{\alpha^\prime}{f_{p_b}(A+A_{ne})}[f_{p_b}(A+A_{ne})-\alpha^\prime]E
\end{align*}
where $E=\log{\frac{f_{p_b}(A+A_{ne})(1-f_{p_b}(A_{ne}))}{f_{p_b}(A_{ne})(1-f_{p_b}(A+A_{ne}))}}$. The upper bound is equal to $$\frac{f_{p_b}(\alpha+A_{ne})}{f_{p_b}(A+A_{ne})}[f_{p_b}(A+A_{ne})-f_{p_b}(\alpha+A_{ne})]\log{\frac{f_{p_b}(A+A_{ne})(1-f_{p_b}(A_{ne}))}{f_{p_b}(A_{ne})(1-f_{p_b}(A+A_{ne}))}}$$ for $\alpha^\prime \leq \frac{f_{p_b}(A+A_{ne})}{2}$ and $\frac{f_{p_b}(A+A_{ne})}{4}\log{\frac{f_{p_b}(A+A_{ne})(1-f_{p_b}(A_{ne}))}{f_{p_b}(A_{ne})(1-f_{p_b}(A+A_{ne}))}}$ otherwise.\\
Now if we consider $f_{p_b}(X+A_{ne})=\frac{X+A_{ne}}{X+A_{ne}+\frac{\kappa}{\gamma}}$ then the upper bound is:
\begin{align*}
&A_{\mathsf{Binomial}}(p(y|x)):=\max_{\substack{p(x):\\ E[X]=\alpha, 0 \leq X \leq \mathsf{A}}}\mathcal{U}(p(x,y))\\\nonumber
&=nN\begin{cases}\frac{f_{p_b}(\alpha+A_{ne})}{f_{p_b}(A+A_{ne})}[f_{p_b}(A+A_{ne})-f_{p_b}(\alpha+A_{ne})]E,&\textit{if},\quad(*)
\\\nonumber
\frac{f_{p_b}(A+A_{ne})}{4}E,&\textit{if},\quad (**)
\end{cases}
\end{align*}
where, $E=\log{\frac{f_{p_b}(A+A_{ne})(1-f_{p_b}(A_{ne}))}{f_{p_b}(A_{ne})(1-f_{p_b}(A+A_{ne}))}}$, $(*)=f_{p_b}(\alpha+A_{ne})< \frac{f_{p_b}(A+A_{ne})}{2}$ and $(**)=f_{p_b}(\alpha+A_{ne})\geq \frac{f_{p_b}(A+A_{ne})}{2}$ .

\section{Proof of lemma \ref{lemma1}}
\label{AppendixProoflemma1}
Let's define:
\begin{align*}
&H^N(p_b)=\\
&-\sum_{y=0}^N {N \choose y}p_b^y(1-p_b)^{N-y} \log({N \choose y}p_b^y(1-p_b)^{N-y}),\\
&p_c=(1-p_b)^{nN}
\end{align*}
Then, the binomial channel transition probabilities by considering binary input, is characterized by:
\begin{align*}
&p(y=0|x=0)=1, \quad p(y=i|x=0)=0,\quad i=1,...,nN, \\
&p(y=i|x=1)={nN \choose i}p_b^i(1-p_b)^{nN-i},\quad  i=1,...,nN,
\end{align*}
Assume $p(x=0)=\alpha$ and $p(x=1)=1-\alpha$. The lower bound on Binomial channel capacity could be derived as below:
\begin{align*}
 &C_{lower} =\max_{\alpha} H(Y) - H(Y|X)\\
 &=\max_{\alpha} H(Y) - \sum_{x=0}^1 H(Y|x)p(x)\\
 &=\max_{\alpha} H(Y) - (1-\alpha) H(Y|x=1)\\
 &=\max_{\alpha} -\sum_{i =1}^{nN}((1-\alpha)p(y=i|x=1))\log{((1-\alpha)p(y=i|x=1))}\\
 & \quad-(\alpha+(1-\alpha)p_c)\log{(\alpha+(1-\alpha)p_c)} - (1-\alpha) H(Y|x=1)\\
 &=\max_{\alpha} -\sum_{i =1}^{nN}((1-\alpha)p(y=i|x=1))\log{((1-\alpha)p(y=i|x=1))}\\
 & \quad-(\alpha+(1-\alpha)p_c)\log{(\alpha+(1-\alpha)p_c)} - (1-\alpha) H^{nN}(p_b)\\
 &= \max_{\alpha} -(1-\alpha)(1-p_c) \log{(1-\alpha)}+(1-\alpha) p_c \log{p_c}\\
 &\quad +(1-\alpha) H^{nN}(p_b)-(\alpha+(1-\alpha)p_c)\log{(\alpha+(1-\alpha)p_c)}\\
 &\quad-(1-\alpha) H^{nN}(p_b)\\
 &=\max_{\alpha} -(1-\alpha)(1-p_c) \log{(1-\alpha)}+(1-\alpha) p_c \log{p_c}\\
 &\quad -(\alpha+(1-\alpha)p_c)\log{(\alpha+(1-\alpha)p_c)}\\
\end{align*}
Taking a derivative with respect to $\alpha$ and setting to zero we obtain:
\begin{align*}
\alpha=\frac{1}{1+\frac{1}{e^{\frac{-p_c \log p_c}{1-p_c}}-p_c}}
\end{align*}
Then
\begin{align*}
C_{lower}=H(\frac{1}{1+e^{g(p_c)}})-\frac{g(p_c)}{1+e^{g(p_c)}}
\end{align*}
where $g(p_c)=\frac{H((1-p_b)^{nN})}{1-(1-p_b)^{nN}}$ and $H(p)=-p\log p -(1-p) \log (1-p)$.
\end{document}